\documentclass[12pt,letterpaper]{article}
\usepackage[margin=1in]{geometry}
\usepackage{multicol,lipsum}
\usepackage{color}
\usepackage{comment}
\usepackage{graphicx}
\usepackage[utf8]{inputenc}
\usepackage{aas_macros}
\usepackage{subfigure}
\usepackage{xcolor}
\usepackage{soul}
\usepackage{graphicx}
\usepackage{floatrow}
\usepackage{amsmath}
\usepackage[T1]{fontenc}
\usepackage{ amssymb }
\usepackage{xspace}
\usepackage{hyperref}
\usepackage{ragged2e}
\usepackage{sidecap} 
\usepackage{authblk}
\usepackage[maxbibnames=3,natbib=true]{biblatex}
\usepackage{sidecap}
\usepackage[symbol, hang,flushmargin]{footmisc}

\def\sgra{Sgr~A$^{\ast}$\xspace}        
\def\m87{M87\xspace}                    
\def\uas{$\mu$as\xspace}                
\def\uv{$(u,v)$\xspace}                 
\def\msun{M$_{\odot}$\xspace}           

\def\lsim{\mathrel{\raise.3ex\hbox{$<$\kern-.75em\lower1ex\hbox{$\sim$}}}}
\def\gsim{\mathrel{\raise.3ex\hbox{$>$\kern-.75em\lower1ex\hbox{$\sim$}}}}

\begin{document}
\thispagestyle{empty}

\begin{center}
    \vspace*{-4cm}
    \makebox[\textwidth]{\includegraphics[width=\paperwidth]{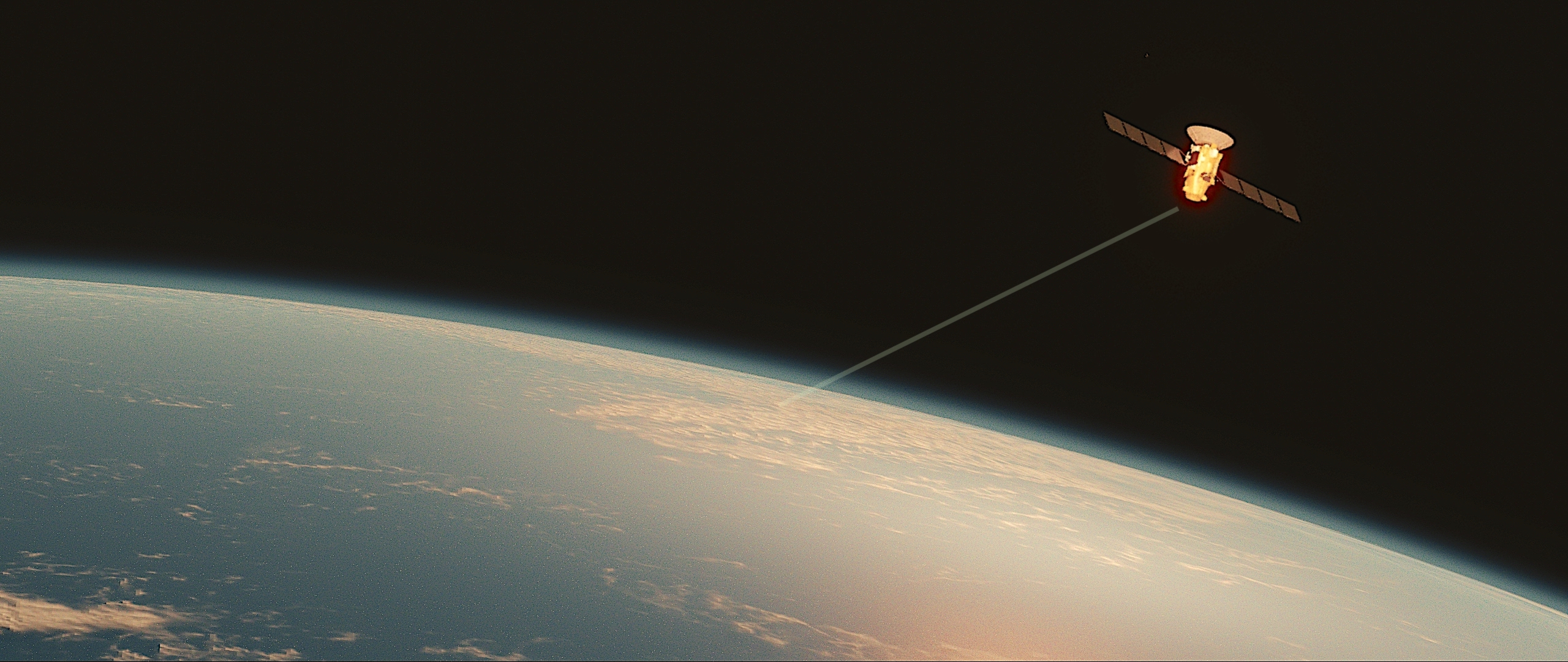}}
\end{center}

\raggedright
\huge
Astro2020 APC White Paper \linebreak

\textbf{Studying black holes on horizon scales with space-VLBI}\linebreak 
\normalsize
\justify

\noindent\textsc{Kari Haworth$^{1,}$\footnote[1]{\href{mailto:kari.haworth@cfa.harvard.edu}{kari.haworth@cfa.harvard.edu}, \href{mjohnson@cfa.harvard.edu}{mjohnson@cfa.harvard.edu}}, 
Michael D. Johnson$^{1,2, \ast}$,
Dominic W. Pesce$^{1,2}$,
Daniel C. M. Palumbo$^{1,2}$,
Lindy Blackburn$^{1,2}$,
Kazunori Akiyama$^{2,3,4,5}$,
Don Boroson$^{6}$,
Katherine L. Bouman$^{7}$,
Joseph R. Farah$^{1,2,8}$,
Vincent L. Fish$^{3}$,
Mareki Honma$^{10,11}$,
Tomohisa Kawashima$^{5}$,
Motoki Kino$^{5,9}$,
Alexander Raymond$^{1,2}$,
Mark Silver$^{6}$,
Jonathan Weintroub$^{1,2}$,
Maciek Wielgus$^{1,2}$,
Sheperd S. Doeleman$^{1,2}$,
Jos\'e L. G\'omez$^{13}$,
Jens Kauffmann$^{3}$,
Garrett K. Keating$^{1}$,
Thomas P. Krichbaum$^{14}$,
Laurent Loinard$^{18,19}$,
Gopal Narayanan$^{12}$,
Akihiro Doi$^{16}$, 
David J. James$^{1,2}$, 
Daniel P.~Marrone$^{15}$,   
Yosuke Mizuno$^{17}$, 
Hiroshi Nagai$^{5}$ }

\begin{multicols}{2}
\noindent\scriptsize{\textit{$^{1}$ Center for Astrophysics $\vert$ Harvard \& Smithsonian, 60 Garden Street, Cambridge, MA 02138, USA \\
$^{2}$ Black Hole Initiative at Harvard University, 20 Garden Street, Cambridge, MA 02138, USA \\
$^{3}$ Massachusetts Institute of Technology, Haystack Observatory, 99 Millstone Road, Westford, MA 01886, USA \\
$^{4}$ National Radio Astronomy Observatory, 520 Edgemont Road, Charlottesville, VA 22903, USA \\
$^{5}$ National Astronomical Observatory of Japan, 2-21-1 Osawa, Mitaka, Tokyo 181-8588, Japan \\
$^{6}$ Massachusetts Institute of Technology, Lincoln Laboratory, 244 Wood St, Lexington, MA 02421\\
$^{7}$ California Institute of Technology, 1200 East California Boulevard, Pasadena, CA 91125, USA \\
$^{8}$ University of Massachusetts Boston, 100 William T, Morrissey Blvd, Boston, MA 02125, USA \\
$^{9}$ Kogakuin University of Technology \& Engineering, Academic Support Center, 2665-1 Nakano, Hachioji, Tokyo 192-0015, Japan \\
$^{10}$ Mizusawa VLBI Observatory, National Astronomical Observatory of Japan, 2-12 Hoshigaoka, Mizusawa, Oshu, Iwate 023-0861, Japan \\
$^{11}$ Department of Astronomical Science, The Graduate University for Advanced Studies (SOKENDAI), 2-21-1 Osawa, Mitaka, Tokyo 181-8588, Japan \\
$^{12}$ Department of Astronomy, University of Massachusetts, 01003, Amherst, MA, USA \\
$^{13}$ Instituto de Astrof\'{\i}sica de Andaluc\'{\i}a-CSIC, Glorieta de la Astronom\'{\i}a s/n, E-18008 Granada, Spain \\
$^{14}$ Max-Planck-Institut f\"ur Radioastronomie, Auf dem H\"ugel 69, D-53121 Bonn, Germany \\
$^{15}$ Steward Observatory and Department of Astronomy, University of Arizona, 933 N. Cherry Ave., Tucson, AZ 85721, USA \\
$^{16}$ The Institute of Space and Astronautical Science, Japan Aerospace Exploration Agency, 3-1-1 Yoshinodai, Chuou-ku, Sagamihara, Kanagawa 252-5210, Japan  \\
$^{17}$ Institut f\"ur Theoretische Physik, Goethe Universit\"at, Max-von-Laue Str. 1, D-60438, Frankfurt am Main, Germany \\
$^{18}$ Instituto de Radioastronom\'{\i}a y Astrof\'{\i}sica, Universidad Nacional Aut\'onoma de M\'exico, Morelia 58089, M\'exico \\
$^{19}$ Instituto de Astronom\'{\i}a, Universidad Nacional Aut\'onoma de M\'exico, CdMx 04510, M\'exico
}}

\end{multicols}
\newpage
\pagenumbering{arabic} 

\section{Introduction}

In 2019, after decades of effort by an international team, the Event Horizon Telescope (EHT) Collaboration presented the first image of a black hole \citep{PaperI,PaperII,PaperIII,PaperIV,PaperV,PaperVI}. The impact of this release, both scientifically and among the public, was extraordinary and felt around the globe. The capability to image black holes on event horizon scales enables entirely new tests of General Relativity (GR) near a black hole and opens a direct window into the astrophysical processes that drive accretion, flaring, and jet genesis.  The EHT image, revealing the supermassive black hole (SMBH) in \m87, was captured using a global very-long-baseline interferometry (VLBI) network operating at 230\,GHz \citep{PaperII}.  Taking the next steps towards precise tests of GR and time-domain studies of accretion flows will require sharper resolution, higher observing frequencies, and faster sampling of interferometric baselines.

The angular resolution of ground-based VLBI is approaching fundamental limits.  Interferometer baseline lengths are currently limited to the diameter of the Earth, imposing a corresponding resolution limit for a ground array of ${\sim}22$\,\uas at an observing frequency of 230\,GHz.  Observations at higher frequencies can improve the resolution but become increasingly challenging because of strong atmospheric absorption and rapid phase variations, severely limiting the number of suitable ground sites and the windows of simultaneous good weather at many global locations.

The extension of the EHT into space with the addition of a single orbiting element would circumvent these limitations and enable a wealth of new scientific possibilities: \vspace{-2mm}

\begin{itemize}
    \item The high time resolution afforded by the rapid \uv-filling will enable reconstructed movies of black hole accretion flows. \vspace{-2mm}
    \item The improved resolution will increase the number of spatially resolvable black hole shadows to dozens, yielding a corresponding number of black hole mass measurements.\vspace{-2mm}
    \item Sharper images will reveal turbulent plasma dynamics, allowing further study of the crucial role magnetic fields play in black hole feeding and in jet launching. \vspace{-2mm}
\end{itemize}

The highest operating frequency of space-VLBI to date is 25 GHz (with RadioAstron), and a number of technical challenges must be overcome to access significantly higher frequencies.  These challenges would be mitigated by anchoring an orbiting element to the highly sensitive elements in the EHT such as ALMA, the LMT, and NOEMA, permitting the use of a modest aperture (of ${\sim}3$-meter size) in space and reducing costs.  In this white paper, we present a conceptual mission design for a first such submillimeter space mission, which we expect to fall within the medium cost category of the Astro2020 survey.  A companion white paper details the concurrent expansion of the EHT ground array.

\begin{figure}[]
    \centering
    \includegraphics[width=0.49 \textwidth]{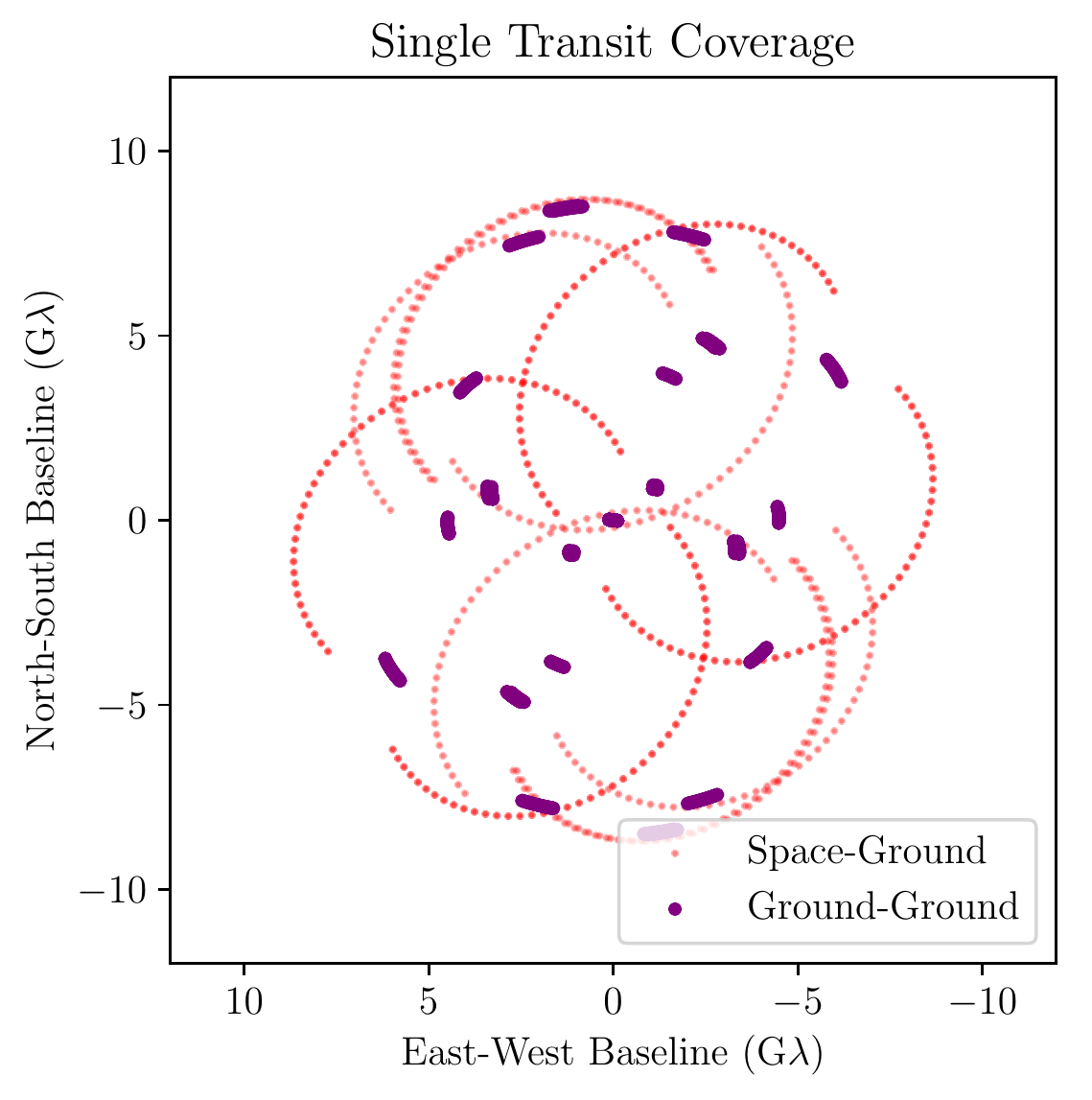}
    \includegraphics[width=0.49 \textwidth]{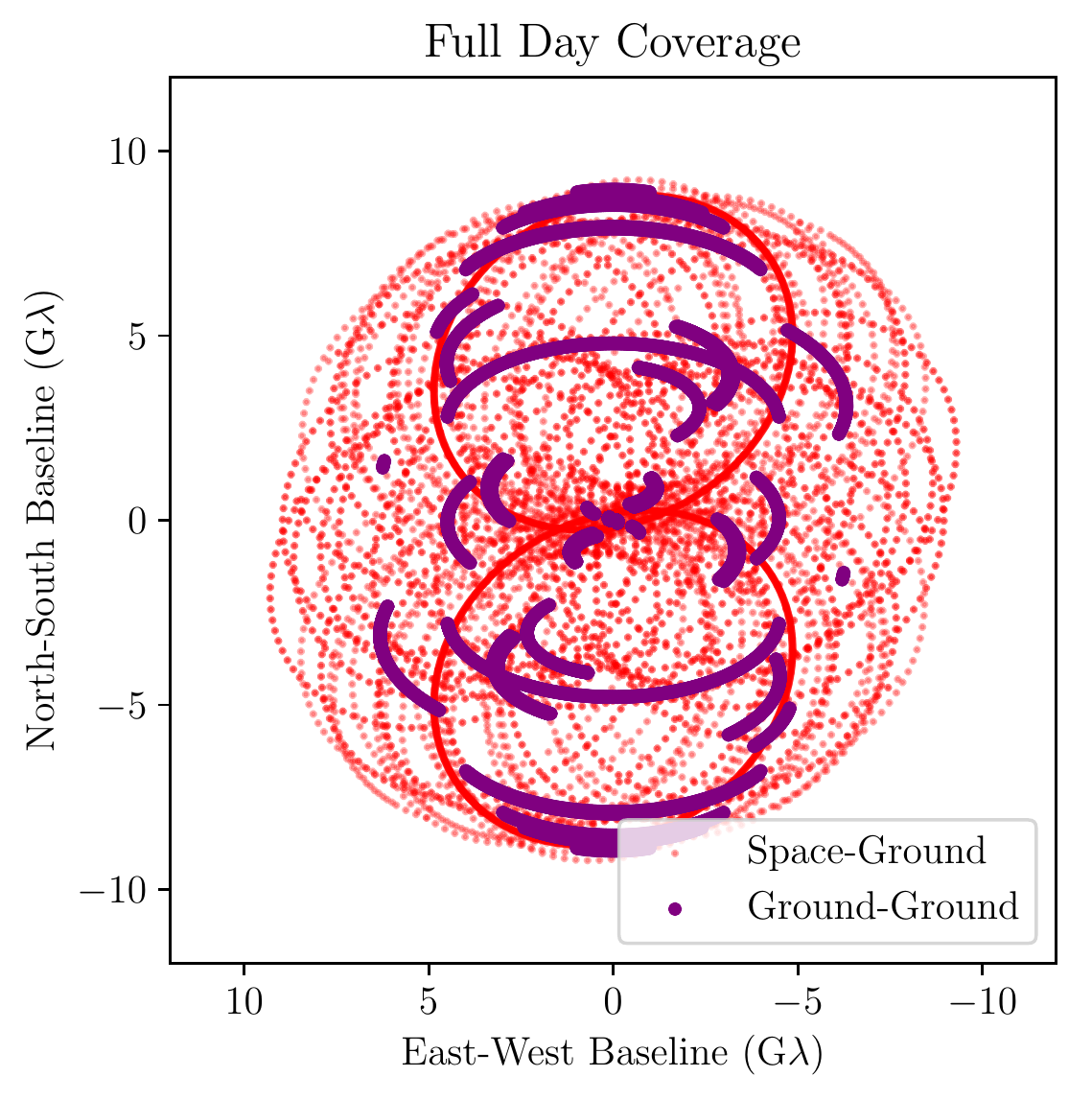}
    \caption{Left: 230 GHz baseline coverage of \sgra of the EHT2020 array with and without a polar LEO over 45 minutes. Right: Same as left, over a full day. In both cases, the addition of the polar LEO dramatically improves the baseline coverage.}
    \label{fig:LEO_coverage}
\end{figure}

\begin{figure}[h]
    \centering
    \includegraphics[width=0.8\textwidth]{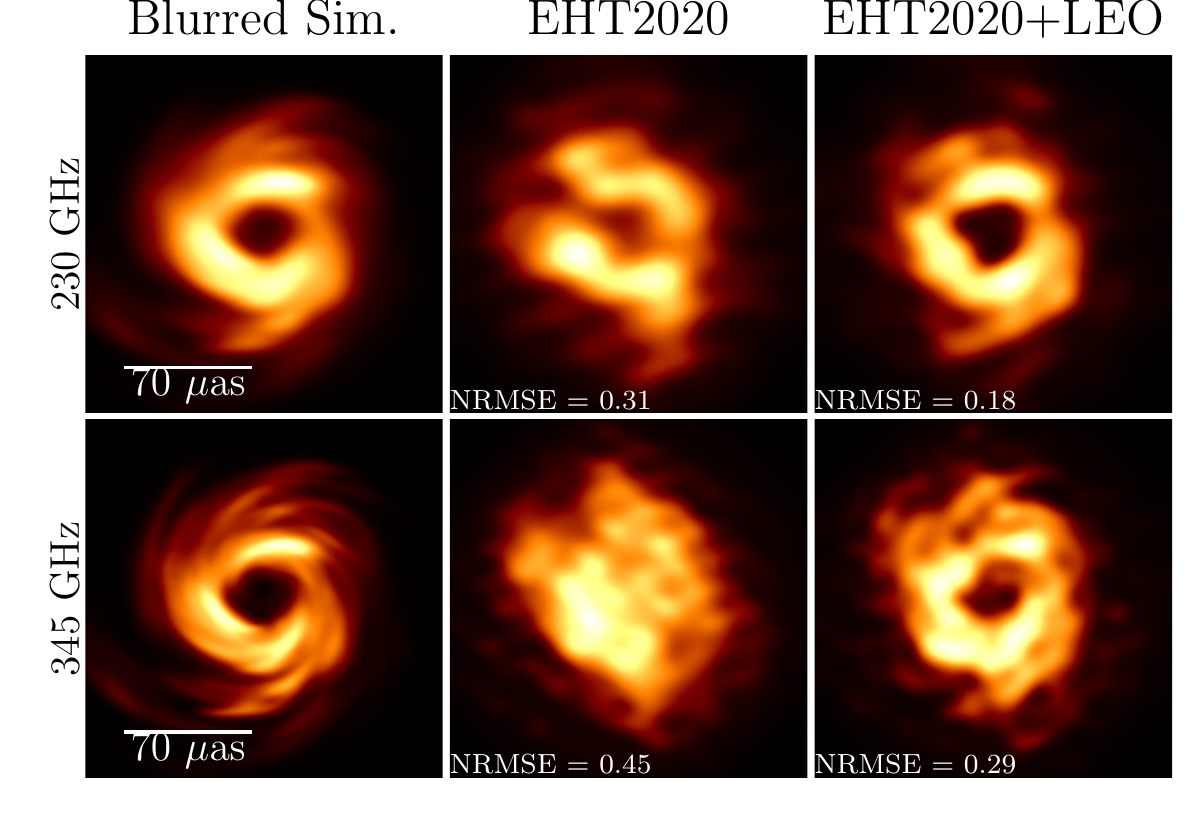}
    \caption{Left column: a ray-traced snapshot from a 40 degree inclined GRMHD simulation of \sgra \citep{Chael_2018_sgra}, blurred by the ensemble average scattering kernel at 230 and 345 GHz \citep{Johnson_2018}. Center, Right columns: reconstructions with the EHT2020 and EHT2020+LEO arrays using 45~minutes of synthetic data.  
    Normalized root-mean-square error (NRMSE) is shown relative to the blurred true image. A single Low Earth Orbiter enables successful reconstruction of the black hole shadow and diffuse plasma features, while the ground arrays have insufficient instantaneous \uv coverage in this short interval.  As discussed in \citep{Palumbo_2019}, sparser sampling at 345 GHz worsens reconstructions that use very brief observations.
    }
    \label{fig:static_imaging}
\end{figure}

\begin{figure}[h]
    \centering
    \includegraphics[width=0.8\textwidth]{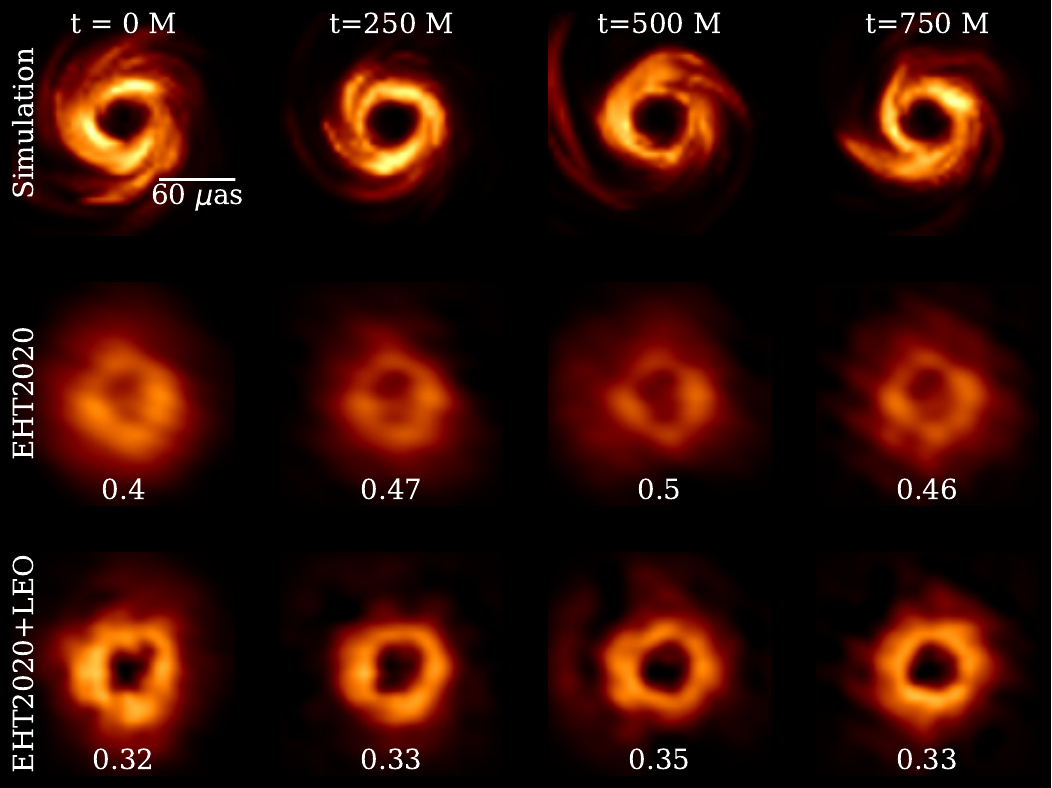}
    \caption{345 GHz Starwarps reconstructions of an ensemble-average scattered \citep{Johnson_2018} GRMHD simulation \citep{Chael_2018_sgra} of Sgr\,A* with the EHT2020 and EHT2020+LEO arrays (matching the simulation in in \autoref{fig:static_imaging}). The normalized root-mean-square error relative to the blurred true image is quoted at the bottom of each panel, showing pixel-wise accuracy for each reconstruction. Frames from the reconstruction are shown every 250\,M, where M is the black hole mass expressed as unit of time via $t_M = G M / c^3$. The improved baseline coverage of the expanded and space-enabled array yield a sharp reconstruction of the black hole shadow, as well as diffuse extended plasma features, while the current ground array produces images dominated by artifacts that do not accurately localize the peak of emission. For additional details, see \citet{Palumbo_2019}. 
    }
    \label{fig:dynamical_imaging}
\end{figure}

\section{Key science goals}

We review the key science drivers for submillimeter space-VLBI.  For additional details, see \citep{Palumbo_2019}, \citep{Fish_2019}, and \citep{Roelofs_2019}.

\subsection{Improved baseline coverage: images and movies of \sgra}

Measuring the shape and diameter of the black hole shadow in \sgra would provide a null hypothesis test of GR \citep{Psaltis_2015} and would yield precise constraints on the surrounding spacetime and the possibility for black hole alternatives.  However, the relatively small mass of this black hole \citep[$4.1 \times 10^6$\,\msun;][]{gravity_redshift_2018} results in correspondingly short dynamical timescales (of order ten minutes) for the system, and the current EHT array lacks sufficient baseline coverage to form images on these timescales.

The addition of a Low-Earth Orbit (LEO) dish to the EHT array would provide sufficient instantaneous \uv-coverage to be sensitive to the ${\lesssim}1$ hour dynamical timescales of \sgra \citep{Palumbo_2019}.  When observing in tandem with ground-based stations, a LEO station accrues coverage in one 45-minute half-orbit comparable to a full night of observation with the expanded EHT array expected for 2020 (EHT2020), as shown in \autoref{fig:LEO_coverage}.  

\autoref{fig:static_imaging} shows static images of a GRMHD simulation of \sgra formed on half-orbit timescales for the EHT2020 with and without a LEO co-observing \citep{Chael_2016,Chael_2018,Chael_2018_sgra}. A LEO orbiter enables reconstruction of the black hole shadow and fine wisps of plasma on single half-orbit timescales.

However, the evolutionary timescale of Sgr A* may not permit static imaging, even over a period as short as 45 minutes. However, the combined \uv-plane filling rate provided by a LEO contribution to the EHT array also enables high-fidelity dynamical imaging of \sgra \citep{Palumbo_2019, Bouman_2018, Johnson_2017}. \autoref{fig:dynamical_imaging} shows dynamical reconstructions of the same GRMHD simulation using the EHT2020 and EHT2020+LEO arrays. The rapid baseline sampling of the orbiter is necessary to recover complex structure in an evolving accretion flow.  

Reconstructed movies of \sgra would elucidate the nature of coherent orbiting features such as ``hotspots''\footnote{By ``hotspot,'' we refer generically to any luminous and compact region on a short--period orbit.} \citep{Gravity_2018} and the origin of the flaring events observed to occur approximately daily across many wavebands \citep{Marrone_2008, YusefZadeh_2009}.  A~large \uv-plane filling fraction will also enable robust modeling in the visibility domain \citep{Broderick_2006,Gravity_2018}, allowing measurements of both the radius and period of very compact orbits.

\begin{figure}[]
    \centering
    \includegraphics[width=0.435 \textwidth]{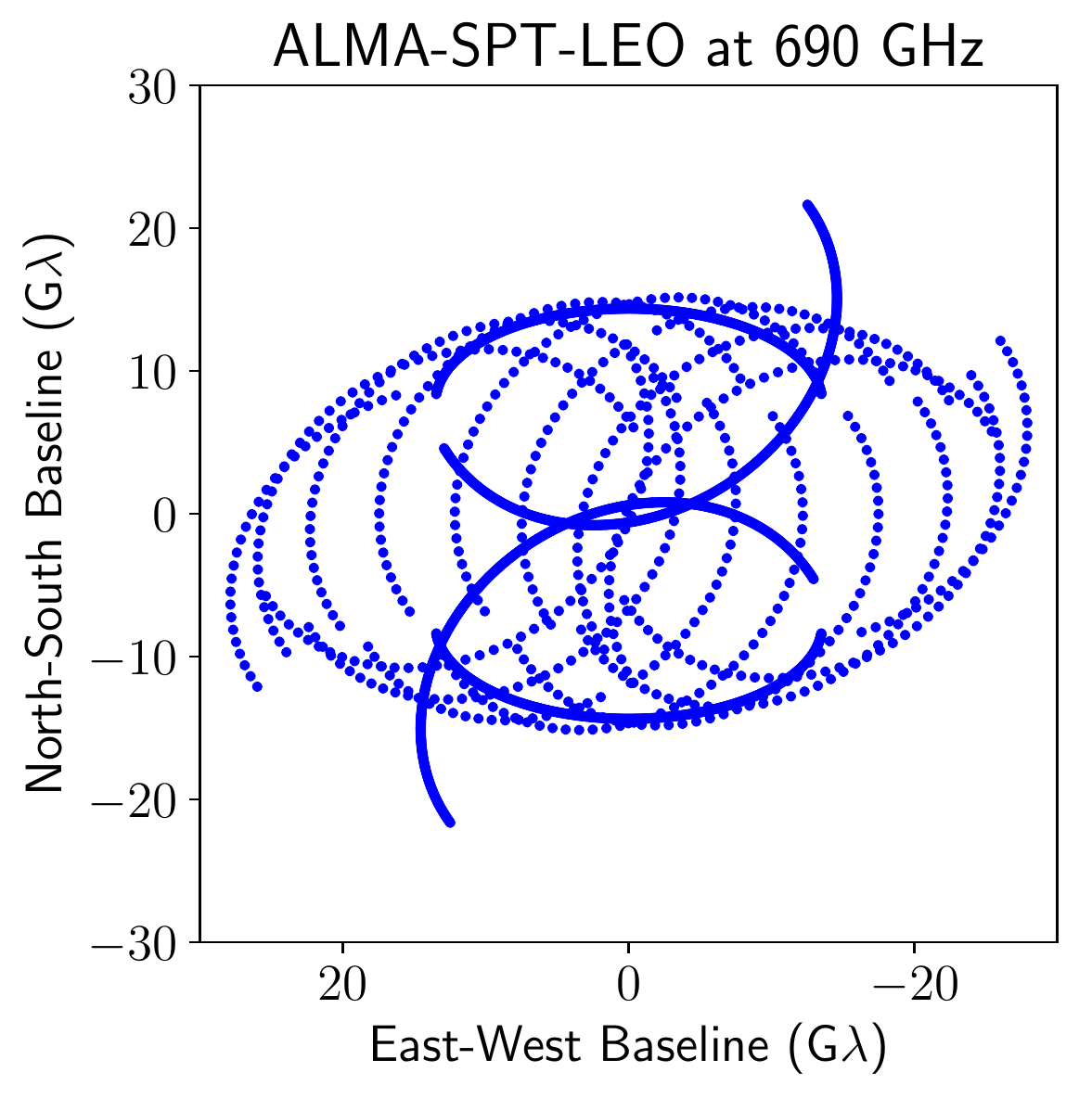}
    \includegraphics[width=0.554 \textwidth]{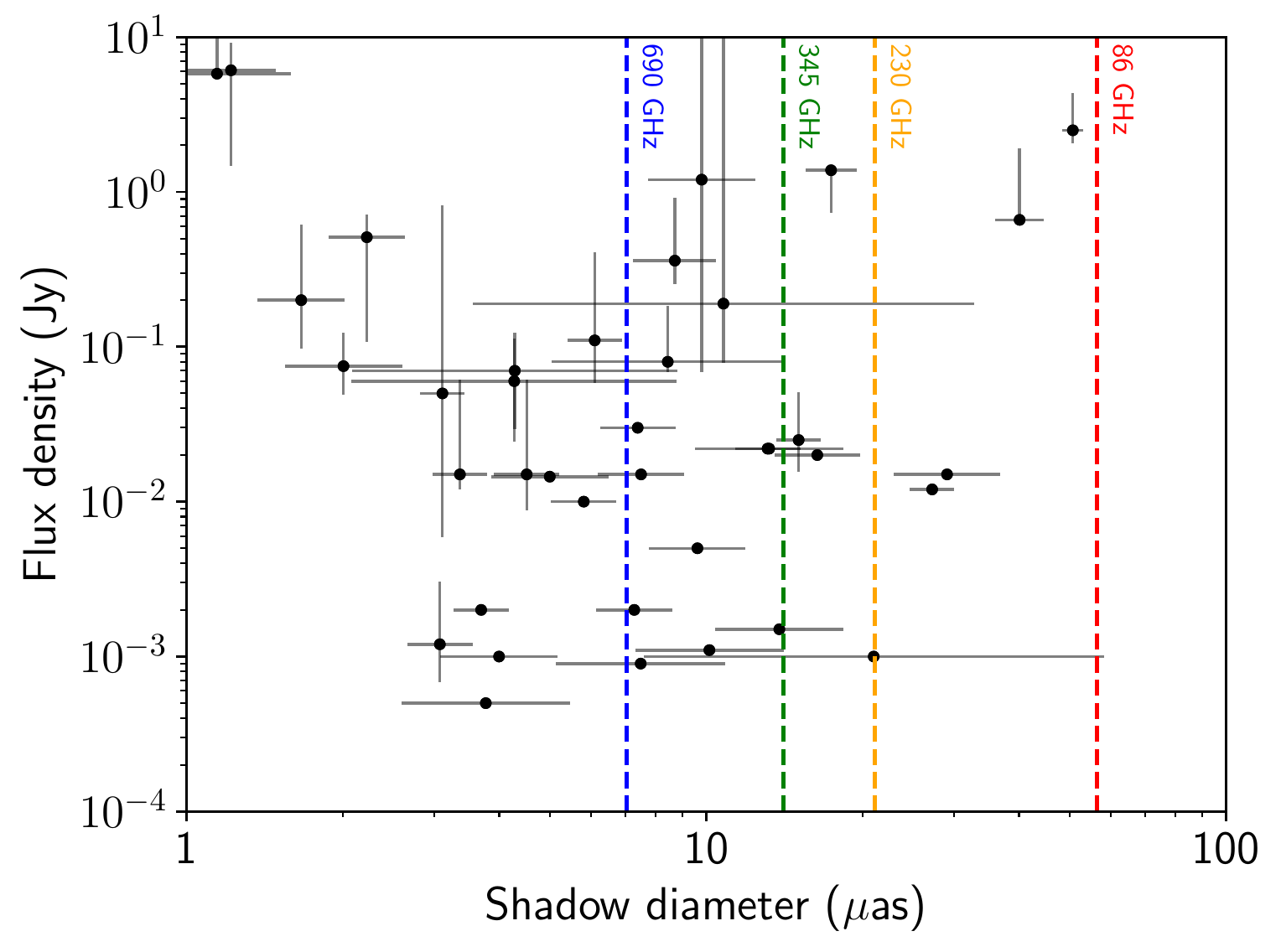}
    \caption{\textit{Left}: \sgra baseline coverage of the ALMA-SPT-LEO subarray at 690 GHz, with points shown every minute over the course of 24 hours. \textit{Right}: 230--690 GHz flux versus black hole shadow size for SMBHs with known masses; fluxes have been taken from the NASA/IPAC Extragalactic Database, and masses were tabulated by \cite{vandenBosch_2016}.  Minimum fringe spacings at different observing frequencies are shown as vertical dashed lines.}
    \label{fig:additional_sources}
\end{figure}

\subsection{Improved resolution: dozens of spatially resolved black holes}

While adding a LEO station to the EHT array would not substantially increase the available physical lengths of baselines, it could improve the angular resolution by extending the observing frequency from 230 GHz up to 690 GHz.  For Earth-diameter baselines, the angular resolution improves from ${\sim}20$\,\uas at 230 GHz to ${\sim}7$\,\uas at 690 GHz.  Even with just two ground stations co-observing with the orbiter, the left panel of \autoref{fig:additional_sources} shows that a LEO station observing at 690 GHz could accumulate a comparable \uv-plane filling fraction to the current (ground-based) EHT array operating at 230 GHz.  Imaging with the 690~GHz coverage alone would thus provide improved resolution, albeit with a limited dynamic range.  In addition, with a LEO station operating at very high frequencies and able to perform observations across a wide range of baseline lengths, we expect to sample the power spectrum of the turbulent accretion flow on very fine spatial scales, for the first time observationally testing our understanding of the magnetorotational instability and angular momentum transport in the inner part of the accretion disks \citep{Balbus_1998,Guan_2009,Walker_2016}.

Finer angular resolution also provides access to additional targets with spatially resolved black hole shadows.  Given a uniform distribution of SMBHs in flat space, we expect the number of sources that can be resolved ($N$) to increase roughly as the cube of the maximum \uv-distance.  At ${\sim}7$\,\uas angular resolution, the number of known SMBHs that are expected to have resolvable black hole shadows will increase from $N \approx 2$ (\sgra and \m87) to $N \gtrsim 20$ (see right panel of \autoref{fig:additional_sources}).\footnote{This number accounts only for those SMBHs with well-measured masses from \cite{vandenBosch_2016}, and so represents a rough lower limit on the number of sources with spatially resolvable black hole shadows.  Synchrotron opacity will preclude horizon-scale observations for some sources, but the opacity decreases with increasing frequency.}  Each spatially resolved shadow provides a corresponding black hole mass-to-distance ratio estimate, a constraint on the black hole spin, and an additional opportunity to study SMBH accretion flow and jet physics on horizon scales.

\section{Technical Overview}
\label{sec::Technical_Overview}

The sensitivity of an interferometric baseline depends on the geometric mean of the two telescope sensitivities, the recorded bandwidth, and the coherent integration time. The first property allows small telescopes (e.g., an orbiter) to form sensitive baselines when paired with a large telescope (e.g., ALMA). The second allows digital enhancements (e.g., wider recorded bandwidths) to offset limitations in telescope sensitivity. The third ties sensitivity to phase stability, which is limited by the atmosphere and reference frequency.

The RMS noise on a VLBI baseline is $\sigma_\text{RMS} = \eta_Q^{-1} \sqrt{(\text{SEFD}_1 \times \text{SEFD}_2) / (2\,\Delta\nu\,T)}$, where $\text{SEFD}_1$ and $\text{SEFD}_1$ are the {\em system-equivalent flux densities} (system noise in units of effective flux above the atmosphere) of each antenna, $\Delta\nu$ and $T$ are the integration bandwidth and time respectively, and $\eta_Q \approx 0.88$ is a loss factor for 2-bit quantization.

For \sgra, the flux density currently seen on the longest EHT baselines is $\sim$100 mJy \citep{Lu_2018}, guiding sensitivity requirements for an orbiter. To maintain a S/N of 4 over 32 GHz (averaging two polarizations) and 5-seconds of integration (sufficient to track rapid atmospheric phase at 230\,GHz), an orbiter--anchor station baseline would require an orbiter SEFD less than $6.2\times10^6$ (ALMA), $1.7\times10^6$ (NOEMA), $3.3\times10^5$ (IRAM30m), $5.8\times10^5$ (LMT), and $1.6\times10^5$ (SMA) to connect successfully to one of the anchor stations in \autoref{tab:anchorstations}. Once phase referenced to the anchor station, the orbiter could also connect to other smaller dishes in a ground array through further coherent averaging.

\begin{SCfigure}[0.9]
  \centering
  \caption{ Sensitivity for a single LEO satellite on a baseline to ALMA as a function of the orbiter SEFD and aggregate bandwidth. Colored regions show expected sensitivity requirements for fringe detections for \sgra\ and M87 (see \S\ref{sec::Technical_Overview} for details). With the current bandwidth of the EHT, a 3.5-meter orbiter would have ample sensitivity for strong detections at 230 and 345\,GHz. At 690\,GHz, detections would require a larger dish or longer integration times, which may be possible with simultaneous multi-frequency observations.\vspace{1.cm}} \label{fig:sensitivity_trades} 
  
  \includegraphics[width=0.5\columnwidth]{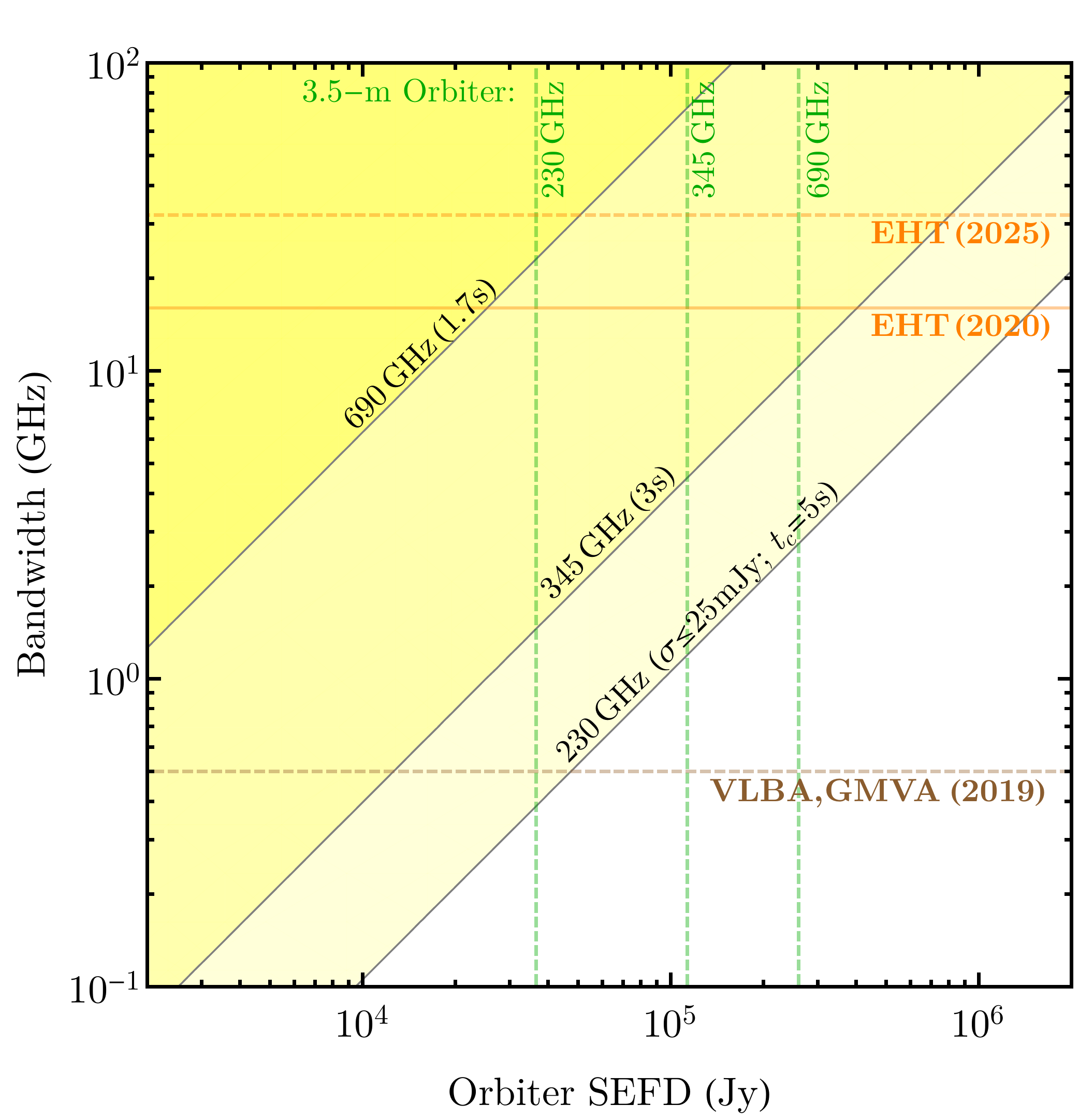}
\end{SCfigure}

Once the required sensitivity for the orbiter is understood, we can conduct trade studies to determine the instrumentation requirements (see Figure~\ref{fig:sensitivity_trades}). Selecting technology that is either presently available or on a near-term development path, the satellite is envisioned to have a 3 - 4 meter dish and will process two polarizations of two 8 GHz bands of data at 230 GHz, 345 GHz, or 690 GHz. Next, we will discuss the details of the instrument subsystems.

\begin{table}[]
    \centering
    \begin{tabular}{lccccccc}
    \hline
    \hline
       & & \textbf{Surface} & \multicolumn{4}{c}{\textbf{SEFD (Jy)}} \\ \cline{4-7}
    \textbf{Telescope} & \textbf{Aperture} & \textbf{rms ($\boldsymbol{\mu}$m)} & 86 GHz & 230 GHz & 345 GHz & 690 GHz \\
    \hline
    ALMA & 54 $\times$ 12\,m & 25 & 33 & 51 & 128 & 1017 \\
    NOEMA & 12 $\times$ 15\,m & 35 & 80 & 188 & 586 & \dots \\
    LMT & 50\,m & 80 & 143 & 563 & 2590 & $\ldots$ \\
    IRAM30m & 30\,m & 55 & 365 & 962 & 2685 & $\ldots$ \\
    SMA & $8 \times 6$\,m & 20 & $\ldots$ & 2030 & 4820  & $\ldots$ \\
    GBT & 100\,m & 300 & 118 & $\ldots$ & $\ldots$ & $\ldots$ \\
    ngVLA (full) & $214 \times 18$\,m & 160 & 7.4 & $\ldots$ & $\ldots$ & $\ldots$ \\
    ngVLA (compact) & $94 \times 18$\,m & 160 & 17 & $\ldots$ & $\ldots$ & $\ldots$ \\
    \hline
    Space & 3.5\,m & 15 & 21600 & 36600 & 113000 & 260000 \\
    \hline
    \end{tabular}
    \caption{Existing and future planned large ground facilities that could serve as anchor stations for a space antenna. For the sensitivity estimates, we have assumed observations at 45 degree elevation, with zenith opacity of 0.05 (ALMA, SMA), 0.09 (NOEMA, IRAM30m, 0.13 (LMT) at 230 GHz, and 0.05 (GBT, and the planned ngVLA \citep{ngvla}) at 86 GHz. Additional beam efficiency of 0.7 was assumed for all sites, except for 0.8 at SMA and a hypothetical 3.5\,m Space orbiter. Receiver temperatures are taken from stations specifications or projections. Receiver temperatures for the Space antenna are based on maturing HEMT technology and are listed in \autoref{sec:receivers}.
    } 
    \label{tab:anchorstations}
\end{table}

\section{Technical Drivers}
\label{sec::tech_drivers}

\begin{figure}[H]
\caption[width=0.9\textwidth]{Space-VLBI instrument for a LEO orbiter with a summary of component development state.} \label{fig:block_diagram}
\includegraphics[trim={0 7.5cm 0 0},clip ,width=1.0\textwidth]{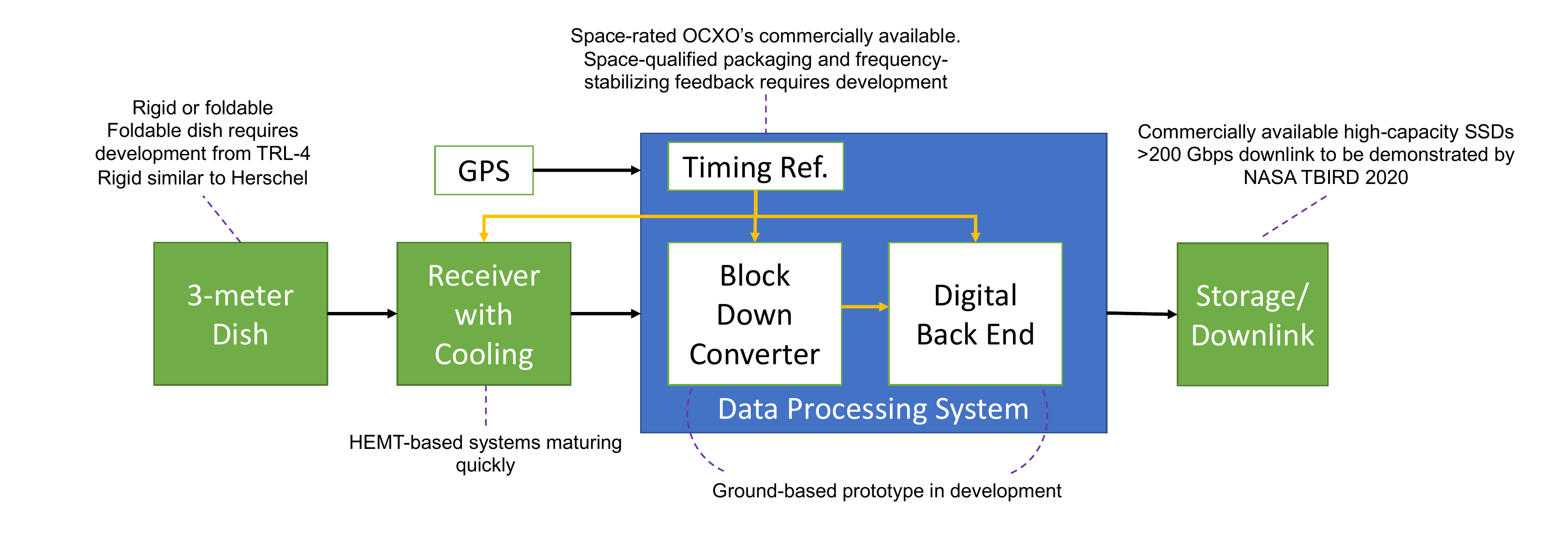}
\end{figure}

Most technical elements of a VLBI satellite are independent of the specific orbital geometry. We now discuss the current status and development trajectories for these common elements, summarized in Figure~\ref{fig:block_diagram}. These include the antenna (\S\ref{sec::antenna}), receiver (\S\ref{sec:receivers}), digital processing system (\S\ref{sec::DPS}), timing reference (\S\ref{sec::timing}), on-board storage and downlink (\S\ref{sec::storage}), and bus (\S\ref{sec::bus}).

\subsection{Antenna}
\label{sec::antenna}
Observing weak sources requires large antenna apertures; however, because anchoring to an Earth-based dish allows for a relatively small reflector in space, the aperture size requirement of 3 - 4 meters is grounded in what's technologically realistic in the near-term. Spacecraft and launch vehicle constraints, not yet determined, define the requirement of size at launch. We review the technology for both a fixed dish, which would require a dedicated launch vehicle, and a deployable dish.

The Herschel satellite flew a fixed 3.5-m antenna with <6 micron surface accuracy \citep{Pilbratt_2010}, sufficient for 230, 345, or 690 GHz operation. A similar concept was developed by Northrup-Grumman Innovation Systems, with a 2-m prototype tested successfully in a balloon launch. Extending to 3.5 meters will require retooling, but the path to build is well understood.

If the launch volume is limited, the precision solid surface reflector segments and supporting structure must be collapsed for launch and deployed once in orbit. Large deployable mesh reflectors operating <40 GHz have significant flight heritage \citep{murphey2009historical}, but deployable reflectors operating at >200 GHz present two new challenges. First, operation at these higher frequencies will require RMS surface accuracy of <20 microns, a level previously unattained in heritage systems. Second, solid surface deployable reflectors will require different, unproven, methods of collapsing the reflector. High-Strain Composite (HSC) precision deployable structures could be used to achieve the precision and alternate deployment approaches with lower size, weight and power than existing technology, and can deploy to within 2.5 microns of their previous position\citep{echter2018recent}. This technology, currently at Technology Readiness Level (TRL) 4, could be matured for flight in 3-5 years. 

\subsection{Receiver}
\label{sec:receivers}

In general, sub-millimeter receivers sensitive enough for VLBI observations require cryogenic cooling.  In space, cooling to 50~K is routine using a mature single-stage Stirling cooler \citep{chattopadhyay2017} and is a reasonable target for a near-term mission. That temperature range prohibits using superconducting-insulator-superconductor (SIS) mixers and instead points to the use of high electron mobility transistor (HEMT) technology.  Quickly-maturing research-grade devices exist for 230, 345, and 690 GHz bands with noise temperatures of 100 \citep{reck2016}, 300 \citep{reck2014,chattopadhyay2017}, and 600~K \citep{reck2014,chattopadhyay2017}, respectively. Additionally, at frequencies above 345~GHz, cooled Schottky diode receivers might be suitable and have flown previously~\citep{melnick2000}.

\subsection{Digital Processing System}
\label{sec::DPS}

The digital processing system (DPS) is required to convert two 8 GHz analog intermediate frequency (IF) signals into digital packets for downlink. The DPS consists of block downconverter (BDC) and a digital backend (DBE). The DBE has two Analog to Digital Converters (ADCs) and a Field Programmable Gate Array (FPGA). Nyquist sampling of 8 GHz requires that the ADC samples at 16 GSps. With four bit samples, this is 128 Gbps on the FPGA input quantized to >64 Gbps (including packet overhead) on the output. While quantization and packetization require minimal processing, the high data throughput means selecting a high-performance FPGA. 

Current ground systems run at 8 Gbps to process two 2-GHz bands \citep{Vertatschitsch_2015}. A new system designed to take advantage of progress made in FPGA and ADC technologies will run at 128 Gbps to process four 8-GHz bands, with a prototype expected within a year. This can be deployed as is to the ground-based system; to use in space will require a redesign of the board for space qualification and to reduce power consumption. The ADC chips and other components have corresponding space versions. Advances have been made in radiation tolerant high-performance FPGAs, with the Xilinx Kintex Ultrascale FPGA providing high data rates in a space-qualified part.

The pathway from the ground-based DPS to a space-ready design is well-understood and is estimated to take three years.

\subsection{Timing reference}
\label{sec::timing}

VLBI requires an extremely stable time reference. The precise specifications required are governed by the observing frequency, the integration time, and the tolerance for phase decoherence. Current ground-based arrays deploy hydrogen masers for this purpose. At 300~GHz observing frequency, a hydrogen maser has adequate phase stability of a part in $10^{15}$ or better, corresponding to coherence losses at 345~GHz less than 5\%, on timescales up to 100~s~\citep{TMS_2017}. 

Hydrogen masers have flight heritage but are large (approx. 35 kg; \citep{Bert2006}) and require significant power (approx. 100 W; \citep{Cacci2006}). High-performance oven-controlled quartz crystal oscillators (OCXO) are small and power efficient, and in laboratory environments, their short-term stability approaches what is needed for VLBI. However, current space-qualified OCXO devices don't meet the requirements, so development of the packaging and frequency-stabilizing feedback is needed if they are to be used as an orbiting VLBI time reference.   

We estimate that OCXO development for space could be completed within two years.

\subsection{Storage/Downlink}
\label{sec::storage}

Packets will be generated at 64 Gbps from the DBE. Assuming a LEO orbit and a 1/2 duty cycle, this results in approximately 22 TB of data per orbit to be handled by a combination of downlink and onboard storage. 

Downlink technology is currently available via laser communication at >100 Gbps. The NASA TBIRD (TeraByte InfraRed Delivery) design is based on commercially-available, highly-integrated 100 Gbps modems, multiplexed into the single telescope using commercially-available wavelength-division-multiplexing fibers, so the downlink speed is configurable to multiple 100's of Gbps at the cost of power and weight. A TBIRD system that will deliver bursts of 200 Gbps to a single ground terminal from a LEO CubeSat is being prepared for a 2020 flight \citep{RobinsonB.S2018TIDT}.  The ground terminals will be small and easily deployable so that several can be fielded, thus increasing the frequency of both seeing the LEO pass over and having a cloud-free line of sight.

For high capacity on-board storage, solid state recorders are available today with 1 TB offered in chewing-gum-sized packages. These are easily multiplexed to much higher data volumes.

\subsection{Bus Considerations: Pointing and positional accuracy}
\label{sec::bus}

VLBI requires that the source be well-localized within the primary beam of each single radio telescope dish. For a 3.5-meter dish diameter, 1.3\,mm observing wavelength, and a pointing requirement to within $\sim$10\% of the beam width, this gives $\delta\theta \lesssim 8"$, which is readily provided through independent optical star-tracking.

Antenna position and velocity must be known in order to coherently average the correlated signal across finite bandwidth and time. Initial searches are conducted with wide search windows in the associated delay, delay-rate, and acceleration, with residual values being used to refine orbit determination, as is currently done for RadioAstron \citep{Zakhvatkin_2018}. Delay windows of $\sim$few $\mu$s ($\sim$km in distance along line-of-sight) are routinely fit in high-bandwidth VLBI post-processing, as well as delay-rate windows of $\sim$few ps/s ($\sim$mm/s). For integrations along Earth orbit, there are often residual acceleration terms which need to be fit as well. This is computationally demanding, but adds a level of complexity that is similar to that required to fit the unknown velocity.

\section{Outlook}
Expansion of the EHT in the coming decade will include both extending the existing ground array, as detailed in a companion white paper, and adding at least one site in space. Though technological considerations differ, these two goals fall under one mission concept: to build on the success of the current EHT and expand spatially resolved studies of black holes. 

A space based antenna in LEO 
would enable routine observations at 345 and 690~GHz.  Operation of the EHT array at higher frequency would permit imaging of dozens of additional black holes.  It would also enable higher precision measurements of the photon ring in targets already observed by EHT, as well as finer-resolution studies of the turbulent plasma surrounding the black hole. Relative to 230 GHz observations, these higher frequencies have lower synchrotron opacity and, for \sgra, sharply reduced interstellar scattering. 

The position of an antenna in LEO changes rapidly compared to the usual changes in VLBI antenna positions driven by the rotation of the Earth.  Baselines to an LEO antenna thus sweep rapidly through the \uv-plane, providing sufficient coverage on short timescales to reconstruct images of \sgra on its dynamical timescale.  The combination of higher observing frequencies and faster baseline sampling---unique to an orbiting VLBI array element---will thus permit time-resolved studies of the Galactic Center and precision tests of general relativity.

While we have not identified any fundamental obstacles in the subsystem technology development paths, all system components require further work on the scale of 2 to 5 years to be ready for a space extension to the EHT. The technical development toward launching an exploratory high-frequency VLBI orbiter in LEO would lay the path for future expansions of the array into medium Earth orbit (MEO; $b_{\text{max}} \approx 1$--3\,D$_{\oplus}$) or geosynchronous orbit (GEO; $b_{\text{max}} \approx 3$\,D$_{\oplus}$), enabling a further factor of $\sim$3 increase in angular resolution, and providing a space VLBI network for multi-wavelength $\mu$as-scale imaging.

\pagebreak

\printbibliography

@inproceedings{murphey2009historical,
  title={Historical perspectives on the development of deployable reflectors},
  author={Murphey, Thomas},
  booktitle={50th AIAA/ASME/ASCE/AHS/ASC Structures, Structural Dynamics, and Materials Conference 17th AIAA/ASME/AHS Adaptive Structures Conference 11th AIAA No},
  pages={2605},
  year={2009}
}

@inproceedings{echter2018recent,
  title={Recent Developments in Precision High Strain Composite Hinges for Deployable Space Telescopes},
  author={Echter, Michael A and Silver, Mark J and D'Elia, Evelyn and Peterson, Michael and Reid, Bryan M},
  booktitle={2018 AIAA Spacecraft Structures Conference},
  pages={0939},
  year={2018}
}

@inproceedings{RobinsonB.S2018TIDT,
pages = {105240V--105240V-6},
volume = {10524},
publisher = {SPIE},
isbn = {9781510615335},
year = {2018},
title = {TeraByte InfraRed Delivery (TBIRD): a demonstration of large-volume direct-to-Earth data transfer from low-Earth orbit},
language = {eng},
author = {Robinson, B. S and Boroson, D. M and Schieler, C. M and Khatri, F. I and Guldner, O and Constantine, S and Shih, T and Burnside, J. W and Bilyeu, B. C and Hakimi, F and Garg, A and Allen, G and Clements, E and Cornwell, D. M},
}

@article{chattopadhyay2017,
author = {Chattopadhyay, Goutam and Reck, Theodore and Kooi, Jacob and Tang, Adrian and Deal, William and Stachnik, Robert A. and Jarnot, Robert and Livesey, Nathaniel},
doi = {10.1109/IRMMW-THz.2017.8066901},
isbn = {9781509060481},
issn = {21622035},
journal = {International Conference on Infrared, Millimeter, and Terahertz Waves, IRMMW-THz},
pages = {1--2},
title = {{A 340 GHz cryogenic amplifier based spectrometer for space based atmospheric science applications}},
year = {2017}
}

@ARTICLE{Balbus_1998,
  author = {{Balbus}, S.~A. and {Hawley}, J.~F.},
    title = "{Instability, turbulence, and enhanced transport in accretion disks}",
  journal = {Reviews of Modern Physics},
     year = 1998,
    month = jan,
  volume = 70,
    pages = {1-53},
      doi = {10.1103/RevModPhys.70.1},
  adsurl = {http://adsabs.harvard.edu/abs/1998RvMP...70....1B}
}

@ARTICLE{YusefZadeh_2009,
  author = {{Yusef-Zadeh}, F. and {Bushouse}, H. and {Wardle}, M. and {Heinke}, C. and 
	{Roberts}, D.~A. and {Dowell}, C.~D. and {Brunthaler}, A. and 
	{Reid}, M.~J. and {Martin}, C.~L. and {Marrone}, D.~P. and {Porquet}, D. and 
	{Grosso}, N. and {Dodds-Eden}, K. and {Bower}, G.~C. and {Wiesemeyer}, H. and 
	{Miyazaki}, A. and {Pal}, S. and {Gillessen}, S. and {Goldwurm}, A. and 
	{Trap}, G. and {Maness}, H.},
    title = "{Simultaneous Multi-Wavelength Observations of Sgr A* During 2007 April 1-11}",
  journal = {\apj},
archivePrefix = "arXiv",
  eprint = {0907.3786},
     year = 2009,
    month = nov,
  volume = 706,
    pages = {348-375},
      doi = {10.1088/0004-637X/706/1/348},
  adsurl = {http://adsabs.harvard.edu/abs/2009ApJ...706..348Y}
}

@ARTICLE{Marrone_2008,
  author = {{Marrone}, D.~P. and {Baganoff}, F.~K. and {Morris}, M.~R. and 
	{Moran}, J.~M. and {Ghez}, A.~M. and {Hornstein}, S.~D. and 
	{Dowell}, C.~D. and {Mu{\~n}oz}, D.~J. and {Bautz}, M.~W. and 
	{Ricker}, G.~R. and {Brandt}, W.~N. and {Garmire}, G.~P. and 
	{Lu}, J.~R. and {Matthews}, K. and {Zhao}, J.-H. and {Rao}, R. and 
	{Bower}, G.~C.},
    title = "{An X-Ray, Infrared, and Submillimeter Flare of Sagittarius A*}",
  journal = {\apj},
archivePrefix = "arXiv",
  eprint = {0712.2877},
     year = 2008,
    month = jul,
  volume = 682,
    pages = {373-383},
      doi = {10.1086/588806},
  adsurl = {http://adsabs.harvard.edu/abs/2008ApJ...682..373M}
}

@article{melnick2000,
	doi = {10.1086/312856},
	url = {https://doi.org/10.1086\%2F312856},
	year = 2000,
	month = {aug},
	publisher = {{IOP} Publishing},
	volume = {539},
	number = {2},
	pages = {L77--L85},
	author = {Gary J. Melnick and John R. Stauffer and Matthew L. N. Ashby and Edwin A. Bergin and Gordon Chin and Neal R. Erickson and Paul F. Goldsmith and Martin Harwit and John E. Howe and Steven C. Kleiner and David G. Koch and David A. Neufeld and Brian M. Patten and Ren{\'{e}} Plume and Rudolf Schieder and Ronald L. Snell and Volker Tolls and Zhong Wang and Gisbert Winnewisser and Yun Fei Zhang},
	title = {The [{ITAL}]Submillimeter Wave Astronomy Satellite[/{ITAL}]: Science Objectives and Instrument Description},
	journal = {The Astrophysical Journal},
}

@article{reck2014,
author = {Reck, Theodore J. and Deal, William and Chattopadhyay, Goutam},
doi = {10.1109/MWSYM.2014.6848250},
isbn = {9781479938698},
issn = {0149645X},
journal = {IEEE MTT-S International Microwave Symposium Digest},
pages = {1--3},
publisher = {IEEE},
title = {{Cryogenic performance of HEMT amplifiers at 340GHz and 670GHz}},
year = {2014}
}

@article{reck2016,
author = {Reck, Theodore and Zemora, Alex and Schlecht, Erich and Dengler, Robert and Deal, William and Chattopadhyay, Goutam},
doi = {10.1109/TTHZ.2015.2506552},
issn = {2156342X},
journal = {IEEE Transactions on Terahertz Science and Technology},
number = {1},
pages = {141--147},
publisher = {IEEE},
title = {{A 230 GHz MMIC-Based Sideband Separating Receiver}},
volume = {6},
year = {2016}
}

@ARTICLE{Broderick_2006,
  author = {{Broderick}, A.~E. and {Loeb}, A.},
    title = "{Imaging optically-thin hotspots near the black hole horizon of Sgr A* at radio and near-infrared wavelengths}",
  journal = {\mnras},
  eprint = {astro-ph/0509237},
     year = 2006,
    month = apr,
  volume = 367,
    pages = {905-916},
      doi = {10.1111/j.1365-2966.2006.10152.x},
  adsurl = {http://adsabs.harvard.edu/abs/2006MNRAS.367..905B}
}

@ARTICLE{Pilbratt_2010,
  author = {{Pilbratt}, G.~L. and {Riedinger}, J.~R. and {Passvogel}, T. and 
	{Crone}, G. and {Doyle}, D. and {Gageur}, U. and {Heras}, A.~M. and 
	{Jewell}, C. and {Metcalfe}, L. and {Ott}, S. and {Schmidt}, M.
	},
    title = "{Herschel Space Observatory. An ESA facility for far-infrared and submillimetre astronomy}",
  journal = {\aap},
archivePrefix = "arXiv",
  eprint = {1005.5331},
 primaryClass = "astro-ph.IM",
     year = 2010,
    month = jul,
  volume = 518,
      eid = {L1},
    pages = {L1},
      doi = {10.1051/0004-6361/201014759},
  adsurl = {https://ui.adsabs.harvard.edu/abs/2010A\%26A...518L...1P}
}

@ARTICLE{Vertatschitsch_2015,
  author = {{Vertatschitsch}, L. and {Primiani}, R. and {Young}, A. and 
	{Weintroub}, J. and {Crew}, G.~B. and {McWhirter}, S.~R. and 
	{Beaudoin}, C. and {Doeleman}, S. and {Blackburn}, L.},
    title = "{R2DBE: A Wideband Digital Backend for the Event Horizon Telescope}",
  journal = {\pasp},
     year = 2015,
    month = dec,
  volume = 127,
    pages = {1226},
      doi = {10.1086/684513},
  adsurl = {http://adsabs.harvard.edu/abs/2015PASP..127.1226V}
}

@ARTICLE{Chael_2016,
  author = {{Chael}, A.~A. and {Johnson}, M.~D. and {Narayan}, R. and {Doeleman}, S.~S. and 
	{Wardle}, J.~F.~C. and {Bouman}, K.~L.},
    title = "{High-resolution Linear Polarimetric Imaging for the Event Horizon Telescope}",
  journal = {\apj},
archivePrefix = "arXiv",
  eprint = {1605.06156},
     year = 2016,
    month = sep,
  volume = 829,
      eid = {11},
    pages = {11},
      doi = {10.3847/0004-637X/829/1/11},
  adsurl = {http://adsabs.harvard.edu/abs/2016ApJ...829...11C}
}

@ARTICLE{Johnson_2017,
  author = {{Johnson}, M.~D. and {Bouman}, K.~L. and {Blackburn}, L. and 
	{Chael}, A.~A. and {Rosen}, J. and {Shiokawa}, H. and {Roelofs}, F. and 
	{Akiyama}, K. and {Fish}, V.~L. and {Doeleman}, S.~S.},
    title = "{Dynamical Imaging with Interferometry}",
  journal = {\apj},
archivePrefix = "arXiv",
  eprint = {1711.01286},
     year = 2017,
    month = dec,
  volume = 850,
      eid = {172},
    pages = {172},
      doi = {10.3847/1538-4357/aa97dd},
  adsurl = {http://cdsads.u-strasbg.fr/abs/2017ApJ...850..172J}
}

@BOOK{TMS_2017,
  author = {{Thompson}, A.~R. and {Moran}, J.~M. and {Swenson}, Jr., G.~W.},
    title = "{Interferometry and Synthesis in Radio Astronomy, 3rd Edition}",
booktitle = {Interferometry and Synthesis in Radio Astronomy, by A.~Richard Thompson, James M.~Moran, and George W.~Swenson, Jr.~3rd ed.~Springer, 2017.},
     year = 2017,
      doi = {10.1007/978-3-319-44431-4},
  adsurl = {http://adsabs.harvard.edu/abs/2017isra.book.....T},
publisher = {Springer}
}

@ARTICLE{Bouman_2018, 
author={{Bouman}, K. and {Johnson}, M. and {Dalca}, A. and {Chael}, A. and {Roelofs}, R. and {Doeleman}. S. and {Freeman}, W.~T.}, 
journal={IEEE Transactions on Computational Imaging}, 
title={Reconstructing Video of Time-Varying Sources from Radio Interferometric Measurements}, 
year={2018}, 
volume={4}, 
number={4}, 
pages={512-527}, 
doi={10.1109/TCI.2018.2838452}, 
ISSN={2333-9403}, 
month={},}

@ARTICLE{Chael_2018,
  author = {{Chael}, A.~A. and {Johnson}, M.~D. and {Bouman}, K.~L. and 
	{Blackburn}, L.~L. and {Akiyama}, K. and {Narayan}, R.},
    title = "{Interferometric Imaging Directly with Closure Phases and Closure Amplitudes}",
  journal = {\apj},
archivePrefix = "arXiv",
  eprint = {1803.07088},
     year = 2018,
    month = apr,
  volume = 857,
      eid = {23},
    pages = {23},
      doi = {10.3847/1538-4357/aab6a8},
  adsurl = {http://adsabs.harvard.edu/abs/2018ApJ...857...23C}
}

@ARTICLE{Chael_2018_sgra,
  author = {{Chael}, A. and {Rowan}, M. and {Narayan}, R. and {Johnson}, M. and 
	{Sironi}, L.},
    title = "{The role of electron heating physics in images and variability of the Galactic Centre black hole Sagittarius A*}",
  journal = {\mnras},
archivePrefix = "arXiv",
  eprint = {1804.06416},
 primaryClass = "astro-ph.HE",
     year = 2018,
    month = aug,
  volume = 478,
    pages = {5209-5229},
      doi = {10.1093/mnras/sty1261},
  adsurl = {https://ui.adsabs.harvard.edu/abs/2018MNRAS.478.5209C}
}

@ARTICLE{gravity_redshift_2018,
      author = {{Gravity Collaboration} and {Abuter}, R. and {Amorim}, A. and
         {Anugu}, N. and {Baub{\"o}ck}, M. and {Benisty}, M. and
         {Berger}, J.~P. and {Blind}, N. and {Bonnet}, H. and {Brandner}, W. and
         {Buron}, A. and {Collin}, C. and {Chapron}, F. and {Cl{\'e}net}, Y. and
         {Coud{\'e} Du Foresto}, V. and {de Zeeuw}, P.~T. and {Deen}, C. and
         {Delplancke-Str{\"o}bele}, F. and {Dembet}, R. and {Dexter}, J. and
         {Duvert}, G. and {Eckart}, A. and {Eisenhauer}, F. and {Finger}, G. and
         {F{\"o}rster Schreiber}, N.~M. and {F{\'e}dou}, P. and {Garcia}, P. and
         {Garcia Lopez}, R. and {Gao}, F. and {Gendron}, E. and {Genzel}, R. and
         {Gillessen}, S. and {Gordo}, P. and {Habibi}, M. and {Haubois}, X. and
         {Haug}, M. and {Hau{\ss}mann}, F. and {Henning}, Th. and {Hippler}, S. and
         {Horrobin}, M. and {Hubert}, Z. and {Hubin}, N. and
         {Jimenez Rosales}, A. and {Jochum}, L. and {Jocou}, K. and
         {Kaufer}, A. and {Kellner}, S. and {Kendrew}, S. and {Kervella}, P. and
         {Kok}, Y. and {Kulas}, M. and {Lacour}, S. and {Lapeyr{\`e}re}, V. and
         {Lazareff}, B. and {Le Bouquin}, J. -B. and {L{\'e}na}, P. and
         {Lippa}, M. and {Lenzen}, R. and {M{\'e}rand}, A. and {M{\"u}ler}, E. and
         {Neumann}, U. and {Ott}, T. and {Palanca}, L. and {Paumard}, T. and
         {Pasquini}, L. and {Perraut}, K. and {Perrin}, G. and {Pfuhl}, O. and
         {Plewa}, P.~M. and {Rabien}, S. and {Ram{\'i}rez}, A. and {Ramos}, J. and
         {Rau}, C. and {Rodr{\'i}guez-Coira}, G. and {Rohloff}, R. -R. and
         {Rousset}, G. and {Sanchez-Bermudez}, J. and {Scheithauer}, S. and
         {Sch{\"o}ller}, M. and {Schuler}, N. and {Spyromilio}, J. and
         {Straub}, O. and {Straubmeier}, C. and {Sturm}, E. and
         {Tacconi}, L.~J. and {Tristram}, K.~R.~W. and {Vincent}, F. and
         {von Fellenberg}, S. and {Wank}, I. and {Waisberg}, I. and
         {Widmann}, F. and {Wieprecht}, E. and {Wiest}, M. and {Wiezorrek}, E. and
         {Woillez}, J. and {Yazici}, S. and {Ziegler}, D. and {Zins}, G.},
        title = "{Detection of the gravitational redshift in the orbit of the star S2 near the Galactic centre massive black hole}",
      journal = {\aap},
         year = "2018",
        month = 7,
      volume = {615},
          eid = {L15},
        pages = {L15},
          doi = {10.1051/0004-6361/201833718},
archivePrefix = {arXiv},
      eprint = {1807.09409},
 primaryClass = {astro-ph.GA},
      adsurl = {https://ui.adsabs.harvard.edu/\#abs/2018A&A...615L..15G}
}

@ARTICLE{Johnson_2018,
  author = {{Johnson}, M.~D. and {Narayan}, R. and {Psaltis}, D. and {Blackburn}, L. and 
	{Kovalev}, Y.~Y. and {Gwinn}, C.~R. and {Zhao}, G.-Y. and {Bower}, G.~C. and 
	{Moran}, J.~M. and {Kino}, M. and {Kramer}, M. and {Akiyama}, K. and 
	{Dexter}, J. and {Broderick}, A.~E. and {Sironi}, L.},
    title = "{The Scattering and Intrinsic Structure of Sagittarius A* at Radio Wavelengths}",
  journal = {\apj},
archivePrefix = "arXiv",
  eprint = {1808.08966},
     year = 2018,
    month = oct,
  volume = 865,
      eid = {104},
    pages = {104},
      doi = {10.3847/1538-4357/aadcff},
  adsurl = {http://adsabs.harvard.edu/abs/2018ApJ...865..104J}
}

@ARTICLE{Lu_2018,
  author = {{Lu}, R.-S. and {Krichbaum}, T.~P. and {Roy}, A.~L. and {Fish}, V.~L. and 
	{Doeleman}, S.~S. and {Johnson}, M.~D. and {Akiyama}, K. and 
	{Psaltis}, D. and {Alef}, W. and {Asada}, K. and {Beaudoin}, C. and 
	{Bertarini}, A. and {Blackburn}, L. and {Blundell}, R. and {Bower}, G.~C. and 
	{Brinkerink}, C. and {Broderick}, A.~E. and {Cappallo}, R. and 
	{Crew}, G.~B. and {Dexter}, J. and {Dexter}, M. and {Falcke}, H. and 
	{Freund}, R. and {Friberg}, P. and {Greer}, C.~H. and {Gurwell}, M.~A. and 
	{Ho}, P.~T.~P. and {Honma}, M. and {Inoue}, M. and {Kim}, J. and 
	{Lamb}, J. and {Lindqvist}, M. and {Macmahon}, D. and {Marrone}, D.~P. and 
	{Mart{\'{i}}-Vidal}, I. and {Menten}, K.~M. and {Moran}, J.~M. and 
	{Nagar}, N.~M. and {Plambeck}, R.~L. and {Primiani}, R.~A. and 
	{Rogers}, A.~E.~E. and {Ros}, E. and {Rottmann}, H. and {SooHoo}, J. and 
	{Spilker}, J. and {Stone}, J. and {Strittmatter}, P. and {Tilanus}, R.~P.~J. and 
	{Titus}, M. and {Vertatschitsch}, L. and {Wagner}, J. and {Weintroub}, J. and 
	{Wright}, M. and {Young}, K.~H. and {Zensus}, J.~A. and {Ziurys}, L.~M.
	},
    title = "{Detection of Intrinsic Source Structure at ${\sim}3$ Schwarzschild Radii with Millimeter VLBI Observations of Sagittarius A*}",
  journal = {\apj},
archivePrefix = "arXiv",
  eprint = {1805.09223},
     year = 2018,
    month = may,
  volume = 859,
      eid = {60},
    pages = {60},
      doi = {10.3847/1538-4357/aabe2e},
  adsurl = {http://adsabs.harvard.edu/abs/2018ApJ...859...60L}
}

@ARTICLE{Gravity_2018,
  author = {{GRAVITY Collaboration} and {Abuter}, R. and {Amorim}, A. and 
	{Baub{\"o}ck}, M. and {Berger}, J.~P. and {Bonnet}, H. and {Brandner}, W. and 
	{Cl{\'e}net}, Y. and {Coud{\'e} Du Foresto}, V. and {de Zeeuw}, P.~T. and 
	{Deen}, C. and {Dexter}, J. and {Duvert}, G. and {Eckart}, A. and 
	{Eisenhauer}, F. and {F{\"o}rster Schreiber}, N.~M. and {Garcia}, P. and 
	{Gao}, F. and {Gendron}, E. and {Genzel}, R. and {Gillessen}, S. and 
	{Guajardo}, P. and {Habibi}, M. and {Haubois}, X. and {Henning}, T. and 
	{Hippler}, S. and {Horrobin}, M. and {Huber}, A. and {Jim{\'e}nez-Rosales}, A. and 
	{Jocou}, L. and {Kervella}, P. and {Lacour}, S. and {Lapeyr{\`e}re}, V. and 
	{Lazareff}, B. and {Le Bouquin}, J.-B. and {L{\'e}na}, P. and 
	{Lippa}, M. and {Ott}, T. and {Panduro}, J. and {Paumard}, T. and 
	{Perraut}, K. and {Perrin}, G. and {Pfuhl}, O. and {Plewa}, P.~M. and 
	{Rabien}, S. and {Rodr{\'{\i}}guez-Coira}, G. and {Rousset}, G. and 
	{Sternberg}, A. and {Straub}, O. and {Straubmeier}, C. and {Sturm}, E. and 
	{Tacconi}, L.~J. and {Vincent}, F. and {von Fellenberg}, S. and 
	{Waisberg}, I. and {Widmann}, F. and {Wieprecht}, E. and {Wiezorrek}, E. and 
	{Woillez}, J. and {Yazici}, S.},
    title = "{Detection of orbital motions near the last stable circular orbit of the massive black hole SgrA*}",
  journal = {\aap},
archivePrefix = "arXiv",
  eprint = {1810.12641},
     year = 2018,
    month = oct,
  volume = 618,
      eid = {L10},
    pages = {L10},
      doi = {10.1051/0004-6361/201834294},
  adsurl = {http://adsabs.harvard.edu/abs/2018A\%26A...618L..10G}
}

@ARTICLE{PaperI,
      author = {{Event Horizon Telescope Collaboration} and {Akiyama}, Kazunori and
         {Alberdi}, Antxon and {Alef}, Walter and {Asada}, Keiichi and
         {Azulay}, Rebecca and {Baczko}, Anne-Kathrin and {Ball}, David and
         {Balokovi{\'c}}, Mislav and {Barrett}, John},
        title = "{First M87 Event Horizon Telescope Results. I. The Shadow of the Supermassive Black Hole}",
      journal = {\apj},
         year = "2019",
        month = "4",
      volume = {875},
      number = {1},
          eid = {L1},
        pages = {L1},
          doi = {10.3847/2041-8213/ab0ec7},
      adsurl = {https://ui.adsabs.harvard.edu/abs/2019ApJ...875L...1E}
}

@ARTICLE{PaperII,
      author = {{Event Horizon Telescope Collaboration} and {Akiyama}, Kazunori and
         {Alberdi}, Antxon and {Alef}, Walter and {Asada}, Keiichi and
         {Azulay}, Rebecca and {Baczko}, Anne-Kathrin and {Ball}, David and
         {Balokovi{\'c}}, Mislav and {Barrett}, John},
        title = "{First M87 Event Horizon Telescope Results. II. Array and Instrumentation}",
      journal = {\apj},
         year = "2019",
        month = "4",
      volume = {875},
      number = {1},
          eid = {L2},
        pages = {L2},
          doi = {10.3847/2041-8213/ab0c96},
      adsurl = {https://ui.adsabs.harvard.edu/abs/2019ApJ...875L...2E}
}

@ARTICLE{PaperIII,
      author = {{Event Horizon Telescope Collaboration} and {Akiyama}, Kazunori and
         {Alberdi}, Antxon and {Alef}, Walter and {Asada}, Keiichi and
         {Azulay}, Rebecca and {Baczko}, Anne-Kathrin and {Ball}, David and
         {Balokovi{\'c}}, Mislav and {Barrett}, John},
        title = "{First M87 Event Horizon Telescope Results. III. Data Processing and Calibration}",
      journal = {\apj},
         year = "2019",
        month = "4",
      volume = {875},
      number = {1},
          eid = {L3},
        pages = {L3},
          doi = {10.3847/2041-8213/ab0c57},
      adsurl = {https://ui.adsabs.harvard.edu/abs/2019ApJ...875L...3E}
}

@ARTICLE{PaperIV,
      author = {{Event Horizon Telescope Collaboration} and {Akiyama}, Kazunori and
         {Alberdi}, Antxon and {Alef}, Walter and {Asada}, Keiichi and
         {Azulay}, Rebecca and {Baczko}, Anne-Kathrin and {Ball}, David and
         {Balokovi{\'c}}, Mislav and {Barrett}, John},
        title = "{First M87 Event Horizon Telescope Results. IV. Imaging the Central Supermassive Black Hole}",
      journal = {\apj},
         year = "2019",
        month = "4",
      volume = {875},
      number = {1},
          eid = {L4},
        pages = {L4},
          doi = {10.3847/2041-8213/ab0e85},
      adsurl = {https://ui.adsabs.harvard.edu/abs/2019ApJ...875L...4E}
}

@ARTICLE{PaperV,
      author = {{Event Horizon Telescope Collaboration} and {Akiyama}, Kazunori and
         {Alberdi}, Antxon and {Alef}, Walter and {Asada}, Keiichi and
         {Azulay}, Rebecca and {Baczko}, Anne-Kathrin and {Ball}, David and
         {Balokovi{\'c}}, Mislav and {Barrett}, John},
        title = "{First M87 Event Horizon Telescope Results. V. Physical Origin of the Asymmetric Ring}",
      journal = {\apj},
         year = "2019",
        month = "4",
      volume = {875},
      number = {1},
          eid = {L5},
        pages = {L5},
          doi = {10.3847/2041-8213/ab0f43},
      adsurl = {https://ui.adsabs.harvard.edu/abs/2019ApJ...875L...5E}
}

@ARTICLE{PaperVI,
      author = {{Event Horizon Telescope Collaboration} and {Akiyama}, Kazunori and
         {Alberdi}, Antxon and {Alef}, Walter and {Asada}, Keiichi and
         {Azulay}, Rebecca and {Baczko}, Anne-Kathrin and {Ball}, David and
         {Balokovi{\'c}}, Mislav and {Barrett}, John},
        title = "{First M87 Event Horizon Telescope Results. VI. The Shadow and Mass of the Central Black Hole}",
      journal = {\apj},
         year = "2019",
        month = "4",
      volume = {875},
      number = {1},
          eid = {L6},
        pages = {L6},
          doi = {10.3847/2041-8213/ab1141},
      adsurl = {https://ui.adsabs.harvard.edu/abs/2019ApJ...875L...6E}
}

@ARTICLE{Roelofs_2019,
      author = {{Roelofs}, Freek and {Falcke}, Heino and {Brinkerink}, Christiaan and
         {Mo{\'s}cibrodzka}, Monika and {Gurvits}, Leonid I. and
         {Martin-Neira}, Manuel and {Kudriashov}, Volodymyr and
         {Klein-Wolt}, Marc and {Tilanus}, Remo and {Kramer}, Michael},
        title = "{Simulations of imaging the event horizon of Sagittarius A* from space}",
      journal = {\aap},
         year = "2019",
        month = "May",
      volume = {625},
          eid = {A124},
        pages = {A124},
          doi = {10.1051/0004-6361/201732423},
archivePrefix = {arXiv},
      eprint = {1904.04934},
 primaryClass = {astro-ph.HE},
      adsurl = {https://ui.adsabs.harvard.edu/abs/2019A&A...625A.124R}
}

@ARTICLE{Palumbo_2019,
  author = {{Palumbo}, D.~C.~M. and {Doeleman}, S.~S. and {Johnson}, M.~D. and 
	{Bouman}, K.~L. and {Chael}, A.~A.},
    title = "{Metrics and Motivations for Earth-Space VLBI: Time-Resolving Sgr A* with the Event Horizon Telescope}",
  journal = {arXiv e-prints},
archivePrefix = "arXiv",
  eprint = {1906.08828},
 primaryClass = "astro-ph.IM",
     year = 2019,
    month = jun,
  adsurl = {https://ui.adsabs.harvard.edu/abs/2019arXiv190608828P}
}

@ARTICLE{Fish_2019,
%   author = {{Fish}, V.L. and {Shea}, M. and {Akiyama}, K.},
%    title = "Imaging Black Holes and Jets with a VLBI Array Including Multiple Space-Based Telescopes",
%  journal = {submitted to Advances in Space Research},
%     year = 2019,
%     note = {}
%}

@ARTICLE{Psaltis_2015,
  author = {{Psaltis}, D. and {{\"O}zel}, F. and {Chan}, C.-K. and {Marrone}, D.~P.
	},
    title = "{A General Relativistic Null Hypothesis Test with Event Horizon Telescope Observations of the Black Hole Shadow in Sgr A*}",
  journal = {\apj},
archivePrefix = "arXiv",
  eprint = {1411.1454},
 primaryClass = "astro-ph.HE",
     year = 2015,
    month = dec,
  volume = 814,
      eid = {115},
    pages = {115},
      doi = {10.1088/0004-637X/814/2/115},
  adsurl = {http://adsabs.harvard.edu/abs/2015ApJ...814..115P}
}

@ARTICLE{Zakhvatkin_2018,
      author = {{Zakhvatkin}, M.~V. and {Andrianov}, A.~S. and {Avdeev}, V. Yu. and
         {Kostenko}, V.~I. and {Kovalev}, Y.~Y. and {Likhachev}, S.~F. and
         {Litovchenko}, I.~D. and {Litvinov}, D.~A. and {Rudnitskiy}, A.~G. and
         {Shchurov}, M.~A.},
        title = "{RadioAstron orbit determination and evaluation of its results using correlation of space-VLBI observations}",
      journal = {arXiv e-prints},
         year = "2018",
        month = 12,
          eid = {arXiv:1812.01623},
        pages = {arXiv:1812.01623},
archivePrefix = {arXiv},
      eprint = {1812.01623},
 primaryClass = {astro-ph.IM},
      adsurl = {https://ui.adsabs.harvard.edu/abs/2018arXiv181201623Z}
}

@ARTICLE{Walker_2016,
      author = {{Walker}, Justin and {Lesur}, Geoffroy and {Boldyrev}, Stanislav},
        title = "{On the nature of magnetic turbulence in rotating, shearing flows}",
      journal = {\mnras},
         year = "2016",
        month = "Mar",
      volume = {457},
      number = {1},
        pages = {L39-L43},
          doi = {10.1093/mnrasl/slv200},
archivePrefix = {arXiv},
      eprint = {1512.03739},
 primaryClass = {astro-ph.SR},
      adsurl = {https://ui.adsabs.harvard.edu/abs/2016MNRAS.457L..39W}
}

@ARTICLE{vandenBosch_2016,
      author = {{van den Bosch}, Remco C.~E.},
        title = "{Unification of the fundamental plane and Super Massive Black Hole Masses}",
      journal = {\apj},
         year = "2016",
        month = "Nov",
      volume = {831},
      number = {2},
          eid = {134},
        pages = {134},
          doi = {10.3847/0004-637X/831/2/134},
archivePrefix = {arXiv},
      eprint = {1606.01246},
 primaryClass = {astro-ph.GA},
      adsurl = {https://ui.adsabs.harvard.edu/abs/2016ApJ...831..134V}
}

@INPROCEEDINGS{Cacci2006,
  author = {{Cacciapuoti}, L.},
    title = "{The ACES Mission: Scientific Objectives and Present Status}",
booktitle = {ESA Special Publication},
     year = 2006,
  series = {ESA Special Publication},
  volume = 621,
    month = jun,
      eid = {164},
    pages = {164},
  adsurl = {https://ui.adsabs.harvard.edu/abs/2006ESASP.621E.164C}
}

@inproceedings{Bert2006,
  title={Development of the active space hydrogen maser for the ACES space experiment of ESA},
  author={{Berthoud}, P. and {Goujon}, D. and {Gritti}, D. and {Jornod}, A. and {Weber}, C. and {D{\"u}rrenberger}, M. and {Schweda}, H. and {Roulet}, M. and {Thieme}, B. and {Baister}, G.},
  booktitle={Proceedings of the 20th European Frequency and Time Forum},
  pages={379--383},
  year={2006},
  organization={IEEE}
}

@ARTICLE{ngvla,
  author = {{Selina}, R.~J. and {Murphy}, E.~J. and {McKinnon}, M. and {Beasley}, A. and 
	{Butler}, B. and {Carilli}, C. and {Clark}, B. and {Durand}, S. and 
	{Erickson}, A. and {Grammer}, W. and {Hiriart}, R. and {Jackson}, J. and 
	{Kent}, B. and {Mason}, B. and {Morgan}, M. and {Ojeda}, O.~Y. and 
	{Rosero}, V. and {Shillue}, W. and {Sturgis}, S. and {Urbain}, D.
	},
    title = "{The ngVLA Reference Design}",
  journal = {Science with a Next Generation Very Large Array },
     year = 2018,
    month = dec,
  volume = 517,
    pages = {15},
  adsurl = {https://ui.adsabs.harvard.edu/abs/2018ASPC..517...15S}
}

@ARTICLE{Guan_2009,
      author = {{Guan}, Xiaoyue and {Gammie}, Charles F. and {Simon}, Jacob B. and
         {Johnson}, Bryan M.},
        title = "{Locality of MHD Turbulence in Isothermal Disks}",
      journal = {\apj},
         year = "2009",
        month = "Apr",
      volume = {694},
      number = {2},
        pages = {1010-1018},
          doi = {10.1088/0004-637X/694/2/1010},
archivePrefix = {arXiv},
      eprint = {0901.0273},
 primaryClass = {astro-ph.GA},
      adsurl = {https://ui.adsabs.harvard.edu/abs/2009ApJ...694.1010G}
}

\end{document}